\documentclass[12pt]{article}
\usepackage{amsthm,amssymb,amsmath}
\usepackage[english]{babel}
 1 
1

\newcommand{\C}{\ensuremath{\mathbb{C}}}

\newcommand{\Z}{\ensuremath{\mathbb{Z}}}
\newcommand{\bc}{{\boldsymbol{c}}}
\newcommand{\balpha}{{\boldsymbol{\alpha}}}
\newcommand{\bbeta}{{\boldsymbol{\beta}}}
\newcommand{\p}{{\boldsymbol{p}}}
\newcommand{\Q}{{\boldsymbol{Q}}}
\newcommand{\bS}{{\boldsymbol{\mathrm{S}}}}

\newcommand{\bt}{{\boldsymbol{t}}}

\newcommand{\bpsi}{{\boldsymbol{\psi}}}

\newcommand{\bu}{{\boldsymbol{u}}}
\newcommand{\ba}{{\boldsymbol{a}}}
\newcommand{\bR}{{\boldsymbol{R}}}
\renewcommand{\d}{\operatorname{d}}
\newtheorem{teh}{Theorem}

\newcommand{\be}{\begin{equation}}
\newcommand{\ee}{\end{equation}}
\begin{document}

\title{\sc Integrable Quasiclassical Deformations of Algebraic
Curves.
\thanks{Partially supported by  DGCYT
project BFM2002-01607 and by the grant COFIN 2002 "Sintesi" }}
\author{B. Konopelchenko $^{1}$ and L. Mart\'{\i}nez Alonso$^{2 }$
\\
\emph{ $^1$ Dipartimento di Fisica, Universit\'a di Lecce and Sezione INFN}
\\ {\em 73100 Lecce, Italy}\\
\emph{$^2$ Departamento de F\'{\i}sica Te\'{o}rica II, Universidad
Complutense}\\ \emph{E28040 Madrid, Spain}}
\date{} \maketitle
\begin{abstract}
A general scheme for determining and studying  integrable
deformations of algebraic curves is presented. The method
is illustrated with the analysis of the hyperelliptic case.
An associated multi-Hamiltonian hierarchy of systems of hydrodynamic type is
characterized.
\end{abstract}

\vspace*{.5cm}

\begin{center}\begin{minipage}{12cm}

\end{minipage}
\end{center}
\newpage

\section{Introduction}

Algebraic curves arise in the study of various problems for nonlinear
differential equations. The theory of finite-gap solutions
of integrable equations and the Whitham averaging theory are
, probably, the most known examples of the relevance of algebraic
curves and their deformations in the context of nonlinear integrable
models (see e. g. \cite{1}-\cite{3}).

In this work we deal with a class of integrable deformations
of algebraic curves $\mathcal{C}$ defined  by polynomial equations
\cite{4}
\begin{equation}\label{1}
F(p,k):=p^N-\sum_{n=0}^{N-1}u_n(k)p^n=0,\quad u_n\in \C[k].
\end{equation}
We investigate
deformations $\mathcal{C}(x,t)$
consistent with the degrees of the polynomials $u_n$
and such that there exists an \emph{action} function
$\bS=\bS(k,x,t)$ verifying
\begin{enumerate}
\item The multiple-valued function $\p=\p(k)$ determined by
\eqref{1} can be expressed as
\[
\p=\bS_x.
\]
\item The function $\bS_t$ represents, like $ \p=\bS_x$,  a meromorphic function on $\mathcal{C}(x,t)$ with
poles only at $k=\infty$.
\end{enumerate}
As a consequence of these conditions  $\p$ obeys an equation
of the form
\begin{equation}\label{1a}
\partial_t \p=\partial_x \Q,
\end{equation}
where $\Q:=\bS_t$ is assumed to be of the form
\[
\Q=\sum_{r=0}^{N-1}a_r(k)\p^r,\quad a_r\in\C[k].
\]

This type of deformations arises in a natural way within the
\emph{quasiclassical} (dispersionless)
limit of integrable systems \cite{5}-\cite{14}. For example, the
Gel'fand-Dikii hierarchies are associated
with the spectral problems
\begin{equation}\label{2}
\Big(\partial_x^N-\sum_{n=1}^{N-2}u_n(x)\,\partial_x^n\Big)\psi-k^N\psi=0,
\end{equation}
and their  quasiclassical limit is
determined by the leading order of the
expansion arising from inserting  the ansatz
\[
\psi=\exp\Big(\frac{S}{\epsilon}\Big),\quad \epsilon\rightarrow 0,
\]
and substituting $\partial_x\rightarrow \epsilon\partial_x$ in \eqref{1a}. Thus,  $p:=S_x$ satisfies
\[
p^N-\sum_{n=0}^{N-1}u_n(x)p^n-k^N=0,
\]
and the Gel-fand-Dikii equations for the coefficients
$u_n$ become a hierarchy of systems of hydrodynamic type.

These deformations
of algebraic curves are present not only in the theory of
reductions of the dispersionless KP (dKP) hierarchy but also
in  more general contexts as the universal
Whitham hierarchy \cite{10}, the hierarchies of rational Lax equations
\cite{11}-\cite{12} or the quasiclassical limits
of integrable systems associated with  \emph{energy-dependent} Schr\"{o}dinger
spectral problems \cite{13}-\cite{14}.

In this paper we follow the suggestions of \cite{13} and propose a general  scheme for
analyzing and classifying
these deformations without relying on any particular type of
dispersionless hierarchy.  We take equations \eqref{1}-\eqref{1a} as our starting point
and express the coefficients $u_n$ of \eqref{1} as elementary
symmetric functions of the branches $p_i$ of $\p$ (Vi\`{e}te
theorem) to formulate the corresponding system of evolution equations
for $u_n$. They turn out to admit a simple general
form in terms of symmetric  functions
of $p_i$ involving the so-called \emph{power sums} functions.
The requirement of consistency
with the polynomial character of the coefficients $u_n$ as functions
of $k$ imposes severe constraints to the curve \eqref{1}.
To deal with the problem of characterizing consistent deformations
we develop a technique based on solving Lenard type relations, which
can be viewed as a quasiclassical version of the fruitful \emph{resolvent
method} \cite{15} of the theory of Lax pairs.
Thus, we provide a class of consistent
deformations determined by systems of hydrodynamic type for the
coefficients $u_n$. Moreover,
they have a natural associated  set of Riemann
invariants. The analysis of the completeness of
these sets of Riemann invariants is found to be related to
the existence of gauge symmetries.

 To illustrate our analysis we study in detail the hyperelliptic case
\[
p^2-v(k)\,p-u(k)=0,\quad  u,\, v\in \C[k].
\]
We present a class of quasiclassical deformations of these curves
determined by a hierarchy of compatible multi-Hamiltonian systems
of hydrodynamic type which is analyzed from the $R$-matrix point of view.
A quantum (dispersionful) counterpart of this hierarchy is also discussed.
The analysis of the deformations of both the general case of \eqref{1} and its reductions
will be presented elsewhere.


\section{Algebraic curves }

We start with some notation conventions to enounce
the results of algebraic geometry \cite{16}-\cite{21}
which are particularly helpful in our analysis.

Let $\mathcal{C}$ denote an algebraic curve determined
by \eqref{1}.
Its associated function $\p=\p(k)$ describes a multiple-valued function
determined  by   $N$ branches $p_i=p_i(k)$ ($i=0,\ldots,N-1$)
satisfying
\begin{equation}\label{bran}
F(p,k)=\prod_{i=0}^{N-1} (p-p_i(k)).
\end{equation}
We denote by $\C((k))$
the field of power series in $k$ with at most a finite number of
terms with positive powers
\[
\sum_{n=-\infty}^N a_n k^n,\quad N\in \Z.
\]
The following general result \cite{16}-\cite{17} will be used in our subsequent analysis:

\begin{teh}({\bf Newton Theorem})

There exists a positive integer $l$ such that the $N$ branches
\begin{equation}\label{br}
p_i(z):=\Big(p_i(k)\Big)\Big |_{k=z^l},
\end{equation}
are elements of  $\C((z))$. In other words,
they are Laurent series of finite order as $z\rightarrow\infty$
\[
p_i(z)=\sum_{n=0}^{N_i} a_n^{(i)}z^n+\sum_{n=1}^{\infty}\frac{b_n^{(i)}}{z^n},
\quad i=1,\ldots,N.
\]
Furthermore, if $F(p,k)$ is irreducible as a polynomial over the
field $\C((k))$ then $l_0=N$ is the least permissible $l$ and the
branches $p_i(z)$ can be labeled so that
\[
p_i(z)=p_0(\epsilon^i z),\quad \epsilon:=\exp \frac{2\pi i}{N}.
\]
\end{teh}

\noindent
{\bf Notation convention}
\emph{
 Henceforth, given an algebraic curve $\mathcal{C}$ we
will denote by $z$ the variable associated with the least
positive integer $l_0$  for which the substitution $k=z^{l_0}$
implies $p_i\in\C((z)),\, \forall i$.
}

\vspace{0.5truecm}
\noindent
{\bf Examples}

The curves
\[
p^2-u(k)=0,\quad u(k):=\sum_{i=0}^m u_i k^i,\quad u_m\neq 0,
\]
have the branches
\[
p_{\pm}:=\sqrt{u(k)}=\sqrt{u_mk^m}\,\Big(1+\mathcal{O}(\frac{1}{k})\Big),
\quad k\rightarrow\infty,
\]
so that $F:=p^2-u(k)$ is an irreducible (reducible) polynomial
 over $\C((k))$ for
odd (even)  $m$ and
\[
k=z,\quad \mbox{for even $m$},\quad k=z^2,\quad
\mbox{for odd $m$}.
\]

\vspace{0.5truecm}

  According to Vi\`{e}te theorem \cite{19} we may write the coefficients
$u_n$ of \eqref{1} in terms of the branches $p_i$ as
\begin{equation}\label{vi}
u_n=(-1)^{N-n-1}\mathrm{s}_{N-n},
\end{equation}
where $\mathrm{s}_k=\mathrm{s}_k(p_0,\ldots,p_{N-1})$ are the elementary symmetric
polynomials
\[
\mathrm{s}_k=\sum_{0\leq i_1<\ldots<i_k\leq N-1}p_{i_1}\cdots p_{i_k}.
\]
In our study we will use also the so-called \emph{power sums}
\cite{19}
\[
\mathcal{P}_k=p_0^k+\cdots+p_{N-1}^k,\quad k\geq 0.
\]
These symmetric functions are polynomials in the elementary symmetric
functions $\mathrm{s}_k$ and, consequently, they can be written as polynomials
in the coefficients $u_n$. In order to obtain these polynomials, one can use
Newton recurrence formulas \cite{19}
\begin{align}\label{nf}
\nonumber \mathcal{P}_k&=k\,u_{N-k}+u_{N-k+1}\,\mathcal{P}_1+
\cdots+u_{N-1}\,\mathcal{P}_{k-1},
\quad 1\leq k\leq N,\\\\
\nonumber \mathcal{P}_k&=u_0\,\mathcal{P}_{k-N}+u_1\,
\mathcal{P}_{k-N+1}+\cdots+u_{N-1}\mathcal{P}_{k-1},\quad k> N,
\end{align}
as well as the explicit determinant expressions \cite{20}
\begin{equation}\label{mf}
\mathcal{P}_k=\left|
\begin{array}{ccccc}
u_{N-1}&1&0&\ldots&0\\
-2u_{N-2}&u_{N-1}&1&\ldots&0\\
\multicolumn{5}{c}{\dotfill}\\
(-1)^{k}(k-1)u_{N-k+1}&\multicolumn{3}{c}{\dotfill}&1\\
(-1)^{k+1}ku_{N-k}&\multicolumn{3}{l}{\dotfill}&u_{N-1}
\end{array}
\right |,
\end{equation}
where it is assumed that $u_N:=-1,\quad u_n:=0,\quad n> N$.
 One has also a similar determinant formula for $u_n$ in terms of $\mathcal{P}_k$
\begin{equation}\label{mf1}
u_n=-\frac{(-1)^{N-n}}{(N-n)!}
\left|
\begin{array}{llllc}
\mathcal{P}_1&1&0&\ldots&0\\
\mathcal{P}_2&\mathcal{P}_1&2&\ldots&0\\
\multicolumn{5}{c}{\dotfill}\\
\mathcal{P}_{N-n-1}&\multicolumn{2}{c}{\dotfill}&\mathcal{P}_1&N-n-2\\
\mathcal{P}_{N-n}&\multicolumn{3}{l}{\dotfill}&\mathcal{P}_1
\end{array}
\right |,
\end{equation}
where $0\leq n\leq N-1$.

\section{Quasiclassical deformations of algebraic curves}

Let  us consider the problem of characterizing families
 $\mathcal{C}(x,t)$ ($x,t\in \C$) of
algebraic curves
\begin{equation}\label{4}
F:=p^N-\sum_{n=0}^{N-1}u_n(k,x,t)p^n=0,\quad u_n\in \C[k].
\end{equation}
which admit a multiple-valued function $\bS=\bS(k,x,t)$ ({\bf action function})
verifying
\[
\p=\bS_x.
\]
Furthermore, we assume that
\[
\Q:=\bS_t,
\]
is, like $\p$,  a meromorphic function on  $\mathcal{C}(x,t)$ which has poles at $k=\infty$ only.
Thus we assume that $\Q$ can be expressed in the form
\begin{equation}\label{6}
\Q=\sum_{r=0}^{N-1}a_r(k)\p^r,\quad a_r\in\C[k].
\end{equation}
The coefficients $a_r$, and consequently the function $\Q$, will be
explicitly dependent on $x$ and $t$ also.

Thus  the deformations of \eqref{4} satisfying these requirements are characterized by
equations of the form
\begin{equation}\label{5}
\partial_t \p=\partial_x\Big(\sum_{r=0}^{N-1}a_r(k)\p^r\Big),
\quad a_r\in\C[k].
\end{equation}

We will refer to the flows of the form \eqref{5} as {\bf quasiclasssical
deformations} of the curve \eqref{4}. They are directly
connected to Lax equations of quasiclassical type. To see this
property notice that in terms of the branches of $\p$, the flow
\eqref{5} reduces to the system
\begin{align}\label{7}
\nonumber \partial_t p_i&=\partial_x Q_i,\quad i=0,\ldots,N-1
\\\\
\nonumber Q_i:&=\sum_{r=0}^{N-1}a_r(k)p_i^r.
\end{align}
These equations imply the existence of  $N$ branches
$\mathrm{S}_i=\mathrm{S}_i(z,\cdot,\cdot)\in \C((z))$ of $\bS$
verifying
\[
\d \mathrm{S}_i=p_i \d x+Q_i \d t+ m_i \d z,\quad m_i:=\frac{\partial \mathrm{S}_i}{\partial z},
\]
where $k=z^{l_0}$. Hence we have
\begin{equation}\label{8}
\d p_i\wedge \d x+\d Q\wedge \d t=\d z_i\wedge \d m_i,
\end{equation}
where $z_i=z(p_i,x,t),\quad i=0,\ldots,N-1$ stand for the functions
obtained by  inverting $p_i=p_i(z,x,t)\; (i=0,\ldots,N-1)$. We now consider the change of variables
\[
(p_i,x,t)\mapsto (z_i,x,t),
\]
and use the standard technique of \cite{6}. Thus
by identifying  the coefficients of $\d p_i\wedge \d x,\; \d p_i\wedge \d t$ and
$\d x\wedge \d t$ in \eqref{8}, we find
\begin{align*}
\frac{\partial z_i}{\partial p_i}\frac{\partial m_i}{\partial x}
-\frac{\partial z_i}{\partial x}\frac{\partial m_i}{\partial p_i}&=1,\\
\frac{\partial z_i}{\partial p_i}\frac{\partial m_i}{\partial t}
-\frac{\partial z_i}{\partial t}\frac{\partial m_i}{\partial p_i}&=
\frac{\partial Q_i}{\partial p_i},\\
\frac{\partial z_i}{\partial x}\frac{\partial m_i}{\partial t}
-\frac{\partial z_i}{\partial t}\frac{\partial m_i}{\partial x}&=
\frac{\partial Q_i}{\partial x},
\end{align*}
so that the functions $z_i$ satisfy the quasiclassical Lax equations
\begin{equation}\label{9}
\frac{\partial z_i}{\partial t}=\frac{\partial Q_i}{\partial p_i}\frac{\partial z_i}{\partial x}
-\frac{\partial Q_i}{\partial x}\frac{\partial z_i}{\partial p_i},
\quad i=0,\ldots,N-1.
\end{equation}

Our next step is to characterize the  evolution  law of the
coefficients $u_n$ induced by \eqref{5}.
From \eqref{7} it follows
\[
\frac{\partial u_n}{\partial t}=\sum_{i,j}
\frac{\partial u_n}{\partial p_i}\,
\partial_x(a_j p_i^j).
\]
At this point it is important to use the following identities
\begin{equation}\label{9a}
\frac{\partial u_n}{\partial p_i}=p_i^{N-n-1}-
\sum_{m=n+1}^{N-1}u_m p_i^{m-n-1},
\end{equation}
which derive from \eqref{1} by differentiating  with respect to
$p_i$ and identifying coefficients of powers of $p$ in the
resulting equation
\[
\frac{F}{p-p_i}=\sum_n \frac{\partial u_n}{\partial p_i}\,p^n.
\]
Thus, one deduces at once that in terms of the variables $u_n$ the
flow \eqref{5} reduces to the system
\begin{equation}\label{10}
\partial_t \bu=J_0\ba,
\end{equation}
where we are denoting
\[
\bu:=\left(
\begin{array}{c}
u_{N-1}\\
\vdots\\
u_{0}
\end{array}
\right),\quad
\ba:=\left(
\begin{array}{c}
a_0\\
\vdots\\
a_{N-1}
\end{array}
\right ),
\]
and $J_0$ is the matrix differential operator
\begin{equation}\label{10a}
J_0:= T^{\top} V^{\top}\partial_x V,
\end{equation}
where
\begin{equation}\label{t}
T:=\left(
\begin{array}{cccc}
1&-u_{N-1}&\cdots&-u_1\\
0&1&\cdots&-u_2\\
\multicolumn{4}{c}{\dotfill}\\
0&\multicolumn{2}{c}{\dotfill}&1
\end{array}
\right),
\end{equation}
and $V$ is the Vandermonde matrix
\begin{equation}\label{v}
V:=\left(
\begin{array}{cccc}
1&p_{N-1}&\cdots&p_{N-1}^{N-1}\\
\multicolumn{4}{c}{\dotfill}\\
\multicolumn{4}{c}{\dotfill}\\
1&p_0&\cdots&p_0^{N-1}
\end{array}
\right).
\end{equation}

The explicit
expression of the operator $J_0$ in terms of the coefficients
$u_n$ follows from the observation that the matrix elements of
$V^{\top}\partial_x V$ can be written as
\begin{equation}\label{11}
\Big(V^{\top}\partial_x V\Big)_{ij}=\mathcal{P}_{i+j}\,\partial_x+
\frac{j}{i+j}\,\mathcal{P}_{i+j,x},\quad 0\leq i,j\leq N-1,
\end{equation}
where $\mathcal{P}_k(p_0,\ldots,p_{N-1})$ are the power sums of the variables $p_i$.
Hence, the system \eqref{10} becomes ($u_N:=-1$)

\begin{equation}\label{12}
\partial_t u_n=-\sum_{r=0}^{N-n-1}\sum_{m=0}^{N-1}u_{n+r+1}
\Big( \mathcal{P}_{m+r}\partial_x+\frac{m}{m+r}\mathcal{P}_{m+r,x}\Big) a_m.
\end{equation}

\noindent
{\bf Example  1}
\vspace{0.3cm}

For $N=2$, the equation for  the curve is
\begin{equation}\label{n2}
F:=p^2-u_1\,p-u_0=0,
\end{equation}
and the first power sums are
\[
\mathcal{P}_1=u_1,\quad \mathcal{P}_2=u_1^2+2\, u_0.
\]
Thus we find
\begin{equation}\label{j2}
J_0 =
\left(
\begin{array}{cc}
2\partial_x& \partial_x(u_1\cdot)\\\\
-u_1\partial_x& 2u_0\partial_x+u_{0,x}
\end{array}
\right)
\end{equation}

\vspace{0.3cm}
\noindent
{\bf Example 2}
\vspace{0.3cm}

For $N=3$
\begin{equation}\label{n3}
F:=p^3-u_2\,p^2-u_1\,p-u_0=0,
\end{equation}
and
\begin{align*}
\mathcal{P}_1&=u_2,\quad \mathcal{P}_2=2\,u_1+u_2^2,\\
\mathcal{P}_3&=3\,u_0+3\,u_1\,u_2+u_2^3,\\
\mathcal{P}_4&=4\,u_0\,u_2+2\,u_1^2+4\,u_1\,u_2^2+u_2^4,
\end{align*}
\begin{equation}\label{j3}
J_0 =
\left(
\begin{array}{ccc}
3\partial_x & u_2\partial_x+u_{2,x}&(2u_1+u_2^2)\partial_x+(2u_1+u_2^2)_x \\\\
-2u_2\,\partial_x& 2u_1\,\partial_x+u_{1,x}&(3u_0+u_1u_2)\partial_x+2u_{0,x}
+2u_1u_{2,x}\\\\
-u_1\partial_x&3u_0\partial_x+u_{0,x}&u_0u_2\partial_x+2u_0\,u_{2,x}
\end{array}
\right)
\end{equation}
\subsection{Consistency conditions and Lenard relations}

Our main aim is to determine expressions for $\ba$ depending on
$z$ and $\bu$ such that \eqref{12} is consistent with the
polynomial dependence of $\bu$ on the variable $k$. That is to
say, if $d_n:=\mbox{degree}(u_n)$ are the degrees of the
coefficients $u_n$ as polynomials in $k$ , then \eqref{12} must
satisfy
\[
\mbox{degree}(J_o \ba)_n\leq d_n,\quad \forall n.
\]


In case of consistency
\eqref{12} will provide a system of hydrodynamic type for the
coefficients of the polynomials $u_n$ and, as a
consequence of \eqref{7}, the coefficients of the expansions of the branches
\[
p_i(z)=\sum_{n=0}^{N_i} h_n^{(i)}(\bu)z^n+\sum_{n=1}^{\infty}\frac{h_n^{(i)}(\bu)}{z^n},
\quad i=1,\ldots,N,\quad k=z^l,
\]
are conserved densities.

Our main strategy for finding consistent  systems is based on using
Lenard type relations

\begin{equation}\label{13}
J_0\bR=0,\quad \bR:=(R_1,\ldots,R_N)^\top,\;\; R_i\in\C((k)),
\end{equation}
and then considering systems of the form
\begin{equation}\label{14}
\bu_t=J_0\ba,\quad \ba:=\bR_+.
\end{equation}
Here $(\,\cdot\,)_+$ and $(\,\cdot\,)_-$ indicate the parts of non-negative and
negative powers in $k$, respectively.
In these cases from the identity
\[
J_0\ba =J_0\bR_+=-J_0\bR_-,
\]
it is clear that  a sufficient condition for the consistency of \eqref{14} is that
\begin{equation}\label{con}
\mbox{degree}\Big(J_0\Big)_{nm}\leq d_n+1,
\end{equation}
for all $n$ and all $m$ such that $a_m=(\bR_+)_m\neq 0$.

If we impose \eqref{con}
for all $0\leq n,m\leq N-1$, we get a sufficient condition for consistency which only depends
on the  curve \eqref{1} and does not refer to the particular solution of the Lenard relation involved in \eqref{14}.

\vspace{0.3cm}
\noindent
{\bf Examples}
\vspace{0.3cm}

From \eqref{j2} it is straightforward to see that for  $N=2$ the
systems of the form \eqref{14} verifying \eqref{con} for all $0\leq n,m\leq 1$ are
characterized by the constraint
\begin{equation}\label{c2}
d_1\leq d_2+1.
\end{equation}

For $N=3$ if we impose \eqref{con} for all $0\leq n,m\leq 2$
then from \eqref{j3} we get the constraints
\begin{align}\label{c3}
\nonumber &d_2\leq 1,\quad d_1\leq d_2+1,\quad d_2\leq d_1+1,\\\\
\nonumber &d_0\leq d_1+1,\quad d_1\leq d_0+1,
\end{align}
which lead to the following thirteen nontrivial
choices for $(d_0,d_1,d_2)$
\begin{align*}
&(1,0,0),\,(0,1,0),\, (1,1,0),\, (2,1,0), \\
&(0,0,1),\, (1,0,1),\,(0,1,1),\,(1,1,1)\\
&(2,1,1),\,(0,2,1),\,(1,2,1),\,(2,2,1),\\
&(3,2,1).
\end{align*}

Examples of flows \eqref{14} in which some components of $\ba$ vanish identically arise
by imposing reduction conditions to the curve \eqref{1}. For instance, let us consider
the curve \cite{13}
\[
p^N-u_0(k)=0,\quad u_0\in\C[k].
\]
The branches of $\p$ are
\begin{equation}\label{13aa}
p_i(k)=\epsilon^ip_0(k) ,\quad p_0(k):=\sqrt[N]{u_0(k)},\quad \epsilon:=\exp\Big(\frac{2\pi}{N}i\Big),
\end{equation}
with $\sqrt[N]{u_0(k)}$ being a given $N$-th root of $u_0(k)$. The systems
\eqref{7} which are compatible with \eqref{13aa} are those of the form
\[
\partial_t p_i=\partial_x\Big(a_1(k)p_i\Big),
\]
so that the corresponding evolution law  for $u_0$ is
\[
\partial_t u_0=J_0\,a_1,\quad J_0:=N\,p_0^{N-1}\partial_x(p_0\cdot)=N\,u_o\,\partial_x+u_{0,x}.
\]
Thus, if $d$ denotes the degree of $u_0(k)$, we get the following solutions of the Lenard relation $J_0R=0$ in $\C((k))$
\[
R_M=\frac{k^{M+\frac{d}{N}}}{p_0(k)}=\frac{k^{M+\frac{d}{N}}}{\sqrt[N]{u_0(k)}},\quad M\geq 0.
\]
They determine the following  infinite set of consistent systems for any degree $d$ of $u_0(k)$
\[
\partial_{t_M}u_0 =(N\,u_o\,\partial_x+u_{0,x})\Big(\frac{k^{M+\frac{d}{N}}}{\sqrt[N]{u_0(k)}}\Big)_+.
\]

\vspace{0.5truecm}

Let us consider now the general case of \eqref{1}. There is a natural class
of solutions of the Lenard relations \eqref{13}. Indeed, from the
expression \eqref{10a} of $J_0$ it is obvious that for
any constant vector $\bc\in\C^N$
\[
\bR:=V^{-1}\,\bc,
\]
is a solution of \eqref{13}. Furthermore, according to \eqref{9a}
\[
V\,T=\Big(\frac{\partial(u_{N-1},\ldots,u_0)}{\partial(p_{N-1},\ldots,p_0)}\Big)^{\top},
\]
so that
\begin{equation}\label{15}
V^{-1}=T\,\Big(\frac{\partial(p_{N-1},\ldots,p_0)}
{\partial(u_{N-1},\ldots,u_0)}\Big)^{\top}.
\end{equation}

In this way,  we have a set of basic solutions $\bR_i=V^{-1}\,\bc_i,\; (\bc_i)_j:=\delta_{ij}$  of the Lenard relations
given by
\begin{equation}\label{16}
\bR_i:=T\, \Big(\frac{\partial p_i}{\partial u_{N-1}}\ldots
\frac{\partial p_i}{\partial u_0}\Big)^{\top}\quad i=0,\ldots, N-1.
\end{equation}
We notice that  direct implicit differentiation of \eqref{1} gives
\begin{equation}\label{16a}
\frac{\partial p_i}{\partial u_j}=\frac{p_i^j}{F_p(p_i)}.
\end{equation}

In summary, systems of the form
\begin{equation}\label{17}
\partial_t \bu = J_0\Big(T\,\nabla C\Big)_{+},
\end{equation}
where
\begin{align*}
&C=\oint_{\gamma} \frac{\d z}{2\pi i}\int \d x\sum_i  f_i(z)\,p_i,\quad
f_i\in \C[z],\\\\
&\nabla C=\Big(\frac{\delta C}
{\delta u_{N-1}}\ldots
\frac{\delta C}{\delta u_0}\Big)^{\top},
\end{align*}
with $\gamma$ being the unit circle in $\C$, are consistent provided
the following conditions are satisfied
\begin{equation}\label{17b}
\mbox{degree}\Big(J_0\Big)_{nm}\leq d_n+1,
\end{equation}
\begin{equation}\label{17bb}
\frac{\delta C}{\delta u_n} \in \C((z^l)),
\end{equation}
for all $0\leq n,m\leq N-1$.

\subsection{Gelfand-Dikii flows}

Let us show how the systems \eqref{17} include the quasiclassical versions of the Gelfand-Dikii
hierarchies \cite{6}. To this end we consider curves  \eqref{1}
of the form
\[
F:=p^N-\sum_{n=1}^{N-1}u_n(x,t)p^n-v_0(x)-z^N=0.
\]
They give rise to a  matrix $T$ which is $z$-independent, so that the corresponding systems
\eqref{17} become
\begin{equation}\label{18}
\partial_t \bu =J\Big(\nabla C\Big)_{+},
\end{equation}
where $J:=J(z,\bu)$ is the symplectic operator
\begin{align}
\nonumber J: =T^{\top} V^{\top}&\partial_x V T
\\\\
\nonumber =\Big(\frac{\partial(u_{N-1},\ldots,u_0)}
{\partial(p_{N-1},\ldots,p_0)}\Big)&\partial_x\,\Big(\frac{\partial(u_{N-1},\ldots,u_0)}
{\partial(p_{N-1},\ldots,p_0)}\Big)^{\top},\\\\
J_{ij}=\sum_{l=0}^i\sum_{m=0}^j u_{N+l-i}\Big( \mathcal{P}_{l+m}&\partial_x+
\frac{m}{l+m}\mathcal{P}_{l+m,x}\Big)\,u_{N+m-j} ,
\end{align}
where we are denoting $u_N:=-1$.

From \eqref{nf} one proves at once that for $N\leq i\leq 2N-1$ the
functions $\mathcal{P}_i$ are linear in $z^N$
\[
\mathcal{P}_i=z^N\, \mathcal{P}_{1,i}+\mathcal{P}_{2,i},
\]
and
\[
T
\left(
\begin{array}{c}
\mathcal{P}_{1,2N-1}
\\\vdots\\
\mathcal{P}_{1,N}
\end{array}
\right)
=
\left(
\begin{array}{c}
\mathcal{P}_{N-1}
\\\vdots\\
\mathcal{P}_{0}
\end{array}
\right)
\]
Hence, as a consequence it follows that $J$ is of the form
\[
J(z,\bu)=z^N\, J_1+J_2.
\]
In particular the operator $J_1$ is given by ($u_N:=-1$)
\[
\Big(J_1\Big)_{ij}=
\left\{\begin{array}{cc}
-\Big((2N-i-j)\,u_{2N-i-j}\,\partial_x+
(N-j)\,u_{2N-i-j,x}\Big) &\mbox{if $i+j\geq N$ }\\\\
0 &\mbox{otherwise}
\end{array}
\right.
\]
The operators  $J_1$ and $J_2$ form a pair of compatible
symplectic operators which describe
the quasiclassical limits of the Gelfand-Dikii
symplectic operators for the standard hierarchies
of scalar Lax pairs \cite{15}.

%

Let us denote by $p_0$
the branch of $\p$ such that
\begin{equation}\label{18a}
p_0=z+\frac{h_0(\bu)}{z}+\ldots \frac{h_n(\bu)}{z^n}+\ldots,\quad z\rightarrow\infty.
\end{equation}
The $N$-th dispersionless Gelfand-Dikii hierarchy can be formulated
as the system of flows
\begin{equation}\label{19}
\frac{\partial z}{\partial t_M}=\frac{\partial Q_M}{\partial p_0}\frac{\partial z_i}{\partial x}
-\frac{\partial Q_M}{\partial x}\frac{\partial z}{\partial p_0},\quad M\geq 1,
\end{equation}
where $z:=z(p_0,x,\bt)$ and
\[
Q_M(p_0,x,\bt):=\Big(z^M\Big)_{\oplus},\quad \bt=(t_1,\ldots,t_n,\ldots).
\]
Here  $(\,\cdot\,)_{\oplus}$ and $(\,\cdot\,)_{\ominus}$
denote the parts of non-negative and negative powers in $p_0$,
respectively. 
It is straightforward to deduce that the Gelfand-Dikii herarchy
\eqref{19} corresponds to the system of flows
\begin{equation}\label{gd}
\partial_{t_M} \bu=J\,\Big(\nabla C_M\Big)_+,
\end{equation}
where the functionals $C_M$ are given by
\[
C_M[\bu]:=\oint_{\gamma}
\frac{\d z}{2\pi i}\int \d x\sum_j(\epsilon^jz)^M \,p_0(\epsilon^j z).
\]
Alternatively, due to the Lenard relation
\[
J\Big(\nabla C_M\Big)_{+}=-
J\Big(\nabla C_M\Big)_{-},
\]
we have
\begin{equation}\label{20}
\partial_{t_M} \bu=
N\, J_2\Big(\nabla  h_{M}\Big)
=-N\,J_1\Big(\nabla h_{M+N}\Big).
\end{equation}

\subsection{Riemann invariants and gauge transformations}

From \eqref{9} it follows that the values
$z_{is}:=z_i(p_s,\bu)$ corresponding to points $p_{i,s}$ at which
\[
\frac{\partial z_i}{\partial p}(p_{i,s},\bu)=0,
\]
are Riemann invariants of the hydrodynamic system \eqref{12}. Therefore
an important question is to know under what conditions these Riemann invariants
are sufficient to integrate \eqref{12}. It turns out that this problem
is closely related to the analysis of the quasiclassical version
of gauge transformations of integrable systems \cite{bd}.

By  a gauge transformation of a consistent system \eqref{5} we
mean a map $\p\rightarrow \p+g,\; g=g(z,x,t)$
\[
p_i\rightarrow p_i+g,\quad g\in\C[z],\; \forall i,
\]
such that the induced transformation on the coefficients $u_n$
\[
u_n\rightarrow \sum_{r=0}^{N-n}
\left(
\begin{array}{c}
n+r\\
r
\end{array}
\right)
\,u_{n+r}\,g^r,
\]
preserves the degrees $d_n$ of $u_n$ as polynomials in $z^l$.


Gauge transformations possess an obvious set of $N-1$ independent
invariants
\[
w_i:=p_i-p_{i+1},\quad i=0,\ldots,N-2,
\]
and a \emph{gauge variable}
\[
\rho:=\frac{1}{N}u_{N-1}=\frac{1}{N}\sum_i p_i,\quad\quad \rho\rightarrow
\rho+g.
\]
Like in the dispersionful case \cite{bd}, we may describe
the dynamical
variables $u_n$ in terms of $N-1$ gauge invariants and the gauge
variable $\rho$.

As far as the Riemann invariants $z_{is}=z_{is}(\bu)$ are concerned
we observe that they satisfy
\[
R(z_{is})=0,
\]
where $R(z)$ is  the discriminant of $F$ (the resultant of the function $F$ and its
derivative $F_p$ with respect to $p$). It is given by \cite{21}
\[
R(z)=(-1)^{N(N-1)/2}\prod_{i>j}(p_i(z)-p_j(z))^2,
\]
which is obviously a gauge invariant. Hence we conclude that
the Riemann invariants $z_{is}=z_{is}(\bu)$ are  gauge invariants.
Therefore, they cannot describe degrees of freedom associated with
gauge variables.

\section{Deformations of Hyperelliptic curves}

The hyperelliptic curves are characterized by quadratic equations in $p$
\begin{equation}\label{21}
F:=p^2-v(k,x)p-u(k,x)=0,
\end{equation}
where
\[
v=\sum_{i=0}^{d_1} v_ik^i,\quad u=\sum_{i=0}^{d_0} u_ik^i.
\]
The  branches of $\p$ are given by
\begin{equation}\label{22}
p_{\pm}=\frac{1}{2}\Big(v\pm\sqrt{v^2+4u}\Big),
\end{equation}
and the map $(p_+,p_-)\mapsto (v,u)$ is
\begin{equation}\label{23}
v=p_++p_-,\quad u=-p_+p_-.
\end{equation}
Let us denote
\[
d:=\mbox{max}(2d_1,d_0).
\]

Notice that $F$ is irreducible (reducible) over  $\C((k))$ for
$d$ odd (even). Thus we set
\begin{equation}\label{23a}
k=z ,\; \mbox{for $d$ even};\quad
k=z^2,\; \mbox{for $d$ odd}.
\end{equation}

According to \eqref{17} we consider equations of the form
\begin{equation}\label{24}
 \partial_t \bu =J_0\Big(T\,
\nabla C\Big)_{+},
\end{equation}
where
\[
\bu:=\left(
\begin{array}{c}
v\\
u
\end{array}
\right),\quad J_0(v,u)=
\left(
\begin{array}{cc}
2\partial_x& \partial_x(v\cdot)\\\\
-v\partial_x& 2u\partial_x+u_x
\end{array}
\right),\quad
T=
\left(
\begin{array}{cc}
1& -v\\\\
0& 1
\end{array}
\right),
\]
for functionals
\[
C=\oint_{\gamma} \frac{\d z}{2\pi i}\int \d x \,(f_+(z)\,p_++f_-(z)p_-),
\]
As we have seen above these systems are consistent provided
\[
d_1\leq d_0+1,
\]
and
\[
\nabla C\in\C((z^2)),\quad \mbox{if $d$ is odd}.
\]
By direct calculation it is straightforward to check  that
\[
T(\nabla C^{(\pm)})=\pm
\left(
\begin{array}{cc}
1& 0\\\\
0& -1
\end{array}
\right)\nabla C^{(\mp)},\quad
C^{(\pm)}:=\oint_{\gamma} \frac{\d z}{2\pi i}\int \d x \,p_{\pm}
\]
Hence, without loss of generality, we can generate the set of flows
\eqref{24} from the hierarchy
\begin{equation}\label{25}
 \partial_{t_N} \bu =J\Big(\nabla C_N\Big)_{+}, \quad N\geq 0,
\end{equation}
where $J=J(u,v)$ is the operator
\begin{equation}\label{26}
J:=\left(
\begin{array}{cc}
-2\partial_x& \partial_x(v\cdot)\\\\
v\partial_x& 2u\partial_x+u_x
\end{array}
\right)
\end{equation}
and
\begin{align}\label{27}
\nonumber C_N[v,u]:&=
\oint_{\gamma} \frac{\d z}{2\pi i}\int \d x \frac{z^N}{2}(p_+-p_-)\\\\
\nonumber &= \oint_{\gamma} \frac{\d z}{2\pi i}\int \d x \frac{z^N}{2}\sqrt{v^2+4u},
\end{align}
with $N$ being  odd when $d$ is odd .

Observe that
\[
\nabla C_N=
\left(
\begin{array}{c}
\frac{\delta C_N}{\delta v}\\\\
\frac{\delta C_N}{\delta u}
\end{array}
\right)_+
=
\left(
\begin{array}{c}
\frac{z^N}{2}\frac{v}{\sqrt{v^2+4u}}\\\\
\frac{z^N}{\sqrt{v^2+4u}}
\end{array}
\right)_+.
\]

\subsection{$R$-matrix theory and
multi-Hamiltonian structure} \vspace{0.4truecm}

The equations \eqref{25} represent a hierarchy of compatible
multi-Hamiltonian systems associated to a $R$-matrix structure. In
order to describe this property we introduce the Lie algebra
$\mathcal{G}$ with elements
\[
\balpha(z,x)=\left(
\begin{array}{c}
\alpha_1(z,x)\\
\alpha_2(z,x)\\
\alpha_3(x)
\end{array}
\right),\quad \alpha_i(\cdot,x)\in \C((z)),\;\;i=1,2,
\]
and commutator defined by
\begin{equation}\label{28}
[\balpha,\bbeta]:=\left(
\begin{array}{c}
\alpha_{1,x}\beta_2-\alpha_2\beta_{1,x}\\
\alpha_{2,x}\beta_2-\alpha_2\beta_{2,x}\\
\int\Big(\alpha_1\beta_{1,x}-\alpha_{1,x}\beta_1\Big)\d x
\end{array}
\right).
\end{equation}
Obviously $\mathcal{G}$ is a central extension of its subalgebra
determined by the constraint $\alpha_3(x)\equiv 0$. The dual space
$\mathcal{G}^*$ of $\mathcal{G}$ is given by the set of elements
of the form
\[
\bu(z,x)=\left(
\begin{array}{c}
v(z,x)\\
u(z,x)\\
c(x)
\end{array}
\right),\quad v(\cdot,x),u(\cdot,x)\in \C((z)),
\]
acting on $\mathcal{G}$ according to
\begin{equation}\label{29}
\langle \bu,\balpha\rangle:=\oint_{\gamma} \frac{\d z}{2\pi i}
 \Big(c\,\alpha_3+\int \d x(
v\,\alpha_1+u\,\alpha_2)\Big).
\end{equation}

It is straightforward to find that the coadjoint action of $\mathcal{G}$
on $\mathcal{G}^*$
\[
\langle ad^*\balpha(\bu),\bbeta\rangle=\langle \bu,[\bbeta,
\balpha]\rangle,
\]
is given by
\begin{equation}\label{30}
ad^*\balpha(\bu)=\left(
\begin{array}{c}
-2\alpha_{1,x}+(v\alpha_2)_x\\\\
v\alpha_{1,x}+2u\alpha_{2,x}+u_x\alpha_2\\\\
0
\end{array}
\right)
=\left(
\begin{array}{c}
J(v,u)
\left(
\begin{array}{c}
\alpha_1\\
\alpha_2
\end{array}
\right)\\
0
\end{array}
\right).
\end{equation}
Hence we conclude that the operator $J(v,u)$ of \eqref{25} is
the symplectic operator  corresponding to
the Lie-Poisson bracket of $\mathcal{G}$.

The functionals (see \eqref{27})
\[
 C_N[\bu]:= \oint_{\gamma} \frac{\d z}{2\pi i}\int \d x
\, \frac{z^N}{2}\sqrt{v^2+4u},
\]
represent Casimir invariants as they satisfy
\begin{equation}\label{31}
J(v,u)\nabla C_N=0.
\end{equation}
Notice that in order to ensure that $\nabla C_N(\bu)\in \mathcal{G}$
we must restrict $C_N$ to the subset of elements
of $\mathcal{G}^*$ for which $\mbox{degree}(v^2+2u)$ at
$z\rightarrow \infty$ is even.

On the other hand , there is an infinite family of $R$-matrices in $\mathcal{G}$ of the
form
\begin{equation}\label{32}
R_m \balpha:=\frac{1}{2}\Big((z^m\balpha)_+-(z^m\balpha)_- \Big),\quad
m\in\Z.
\end{equation}
They define a corresponding family of Lie-algebra structures in
$\mathcal{G}$ given by
\begin{equation}\label{33}
[\balpha,\bpsi]_m=[R_m\balpha,\bpsi]+[\balpha,R_m\bpsi].
\end{equation}

According to the general theory \cite{19}, the Casimir invariants of $(\mathcal{G},[\,,\,])$
are in involution
\[
\{C_N,C_M\}_m=0,\quad N,M\geq 0,
\]
with respect to all the
Poisson-Lie bracket structures $(\mathcal{G},[\,,\,]_m),\,m\in\Z$.

Therefore for each $m\in\Z$, there is an associated family of compatible
Hamiltonian systems
\begin{align}\label{34}
\nonumber \partial_{t_{m,N}}\bu=&\frac{1}{2}\Big(ad^*(R_m \nabla C_N(\bu))\Big)(\bu)
=\Big(ad^*(z^m\nabla C_N(\bu))_+\Big)(\bu)\\\\
\nonumber &=J(v,u)(z^m\nabla C_N(\bu))_+=-J(v,u)(z^m\nabla C_N(\bu))_-.
\end{align}
These Hamiltonian flows are defined on the invariant submanifold
$\mathcal{D}$  given by the elements $\bu$ of $\mathcal{G}^*$ such that
\begin{equation}\label{34a}
d_1\leq d_0+1,\quad d=\mbox{max}(2\,d_1,d_0)=\mbox{even},
\end{equation}
with $d_0$ and $d_1$ being the degrees of $u$ and $v$ as $z\rightarrow
\infty$, respectively. Moreover,  given positive integers $d_0$
and $d_1$ verifying \eqref{34a} then the submanifols
$\mathcal{D}_{d_0,d_1}$
of $\mathcal{D}$ with elements of the form
\begin{equation}\label{35}
v=\sum_{n=0}^{d_{1}} v_n z^{n},\quad u=\sum_{n=0}^{d_{0}} u_n z^{n},
\end{equation}
are left invariant under the flows \eqref{34}

Therefore we conclude that under conditions \eqref{34a} the
flows \eqref{25} form a hierarchy of
compatible Hamiltonian systems for $(v,u)$.

Furthermore, as a consequence of the identities
\[
\nabla C_{N+m}(\bu)=z^m\nabla C_N,
\]
it follows that the flows \eqref{34} have an infinite number of
compatible Hamiltonian formulations. On  the other hand, one finds
that the coadjoint action of $(\mathcal{G},[\,,\,]_m)$ on
$\mathcal{G}^*$ reads
\[
2\,ad^*_m\alpha(\bu)=J(v,u)(R_m\alpha)-z^mR(J(v,u)\alpha).
\]
In particular, \eqref{34a} implies
\begin{align*}
ad^*_m\alpha(\bu)&=J(v,u)(z^m\alpha_-)_+-z^m(J(v,u)\alpha_-)_+\\
&=z^m(J(v,u)\alpha_-)_--J(v,u)(z^m\alpha_-)_-,
\end{align*}
and it means that the images of $v$ and $u$ under
$ad^*_m\alpha$ are polynomials in $z$ with degrees
\[
\tilde{d}_1=\mbox{max}(m-1,d_1-1),\quad
\tilde{d}_2=\mbox{max}(m-1,d_1-1,d_0-1).
\]
Therefore,  for $0\leq m\leq d_1$ the subsets \eqref{35}  determine Poisson
submanifolds of $\mathcal{G}^*$. In this way
we conclude that under conditions \eqref{34a}
 the equations
 \eqref{25} not only form a hierarchy of
compatible Hamiltonian systems for $(v,u)$ but also have $d_1+1$
different Hamiltonian structures.

The same result holds for \eqref{25} if  $d=\mbox{max}(2d_1,d_0)$
is an odd integer. To prove it one applies the above analysis
to the submanifols
$\widetilde{\mathcal{D}}_{d_0,d_1}$
of $\mathcal{D}$ with elements of the form
\begin{equation}\label{35a}
v=\sum_{n=0}^{d_{1}} v_n z^{2\,n},\quad u=\sum_{n=0}^{d_{0}} u_n z^{2\,n}.
\end{equation}

There are many interesting reductions of \eqref{25}. For example
a compatible constraint is $v\equiv 0$. That is to say
\[
F:=p^2-u(k)=0,
\]
so that $p_+=-p_-$ and \eqref{25} becomes
\[
\partial_t u=J\Big(\frac{\delta C_N}{\delta u}\Big)_+,
\]
where
\begin{align*}
J:&=J(u)=2u\partial_x+u_x,\\ \\
 C_N[u]:&=
\oint_{\gamma} \frac{\d z}{2\pi i}\int \d x\, z^N p_+=
\oint_{\gamma} \frac{\d z}{2\pi i}\int \d x\, z^N\sqrt{u}.
\end{align*}
It can be seen that for this case our analysis lead to the
Hamiltonian description obtained from the geometrical approach in \cite{14}.
\subsection{Riemann invariants and gauge transformations}

The discriminant of $F=p^2-v\,p-u$ is given by
\[
R(z)=-(v^2+4\,u),
\]
so that we can formulate the flows \eqref{25} in terms of the
gauge invariant variable $w:=v^2+4\, u$ and the gauge variable
$\rho:=v/2$. Thus, from \eqref{26}-\eqref{27} one finds
\begin{align}
\nonumber \partial_{t_N} w=&(2\,w\partial_x+w_x)
\Big(\frac{z^N}{\sqrt{w}}\Big)_+,\\\\
\nonumber \partial_{t_N}\rho&=-\partial_x\Big(\Big(\frac{z^N}{\sqrt{w}}\Big)_-\rho\Big)_+.
\end{align}
Observe that the variable $w$ evolves independently of $\rho$ and
that the equation for $\rho_t$  is linear in $\rho$. This suggests
the following scheme for the integration of \eqref{25}: We integrate
first the equation for $w$ in terms of Riemann invariants and then
we solve the linear equation for $\rho$.

\vspace{0.3cm}
\noindent
{\bf Example}
\vspace{0.3cm}

The $t_1$-flow for
\[
F:=p^2-(z+v_0)\,p-(z^2+z\,u_1+u_0)=0,
\]
is
\begin{align}\label{e}
\nonumber \partial_t v_0=&\frac{1}{5\sqrt{5}}(v_{0,x}+2\,u_{1,x}),\\
\partial_t u_1=&\frac{1}{5\sqrt{5}}(2\, v_{0,x}+4\,u_{1,x}),\\
\nonumber \partial_t u_0=&\frac{1}{5\sqrt{5}}(2\,v_0v_{0,x}-v_0\,u_{1,x}+u_{0,x}).
\end{align}
In this case, we have the gauge invariants
\begin{align*}
w:&=v^2+4\,u=5z^2+2z\,w_1+w_2,\\
w_1&=v_0+2\,u_1,\quad w_2=v_0^2+4\,u_0.
\end{align*}
They are also Riemann invariants as they verify
\[
\partial_t w_i=\frac{1}{\sqrt{5}}\partial_x w_i,\quad i=1,2.
\]
Moreover, the gauge variable $\rho_0:=v_0/2$ evolves according to
\[
\partial_t \rho_0=\frac{1}{10\sqrt{5}}\partial_x w_1.
\]
In this way, the integration of \eqref{e} reduces to elementary
operations.

\subsection{Dispersionful counterpart}

There is a dispersionfull version  of the hierarchy of flows
\eqref{25} which can be described in terms of the energy-dependent
spectral problem
\begin{equation}\label{36}
L\,\psi:=\partial_{xx}\psi-v(k,x)\psi_x-u(k,x)\psi=0,
\end{equation}
where
\[
v=\sum_{i=0}^{d_1} v_ik^i,\quad u=\sum_{i=0}^{d_0} u_ik^i.
\]
Indeed,the compatibility between \eqref{36} and flows of the form
\[
\partial_t \psi=a\,\psi+b\, \psi_x,\quad a,b\in\C[z],
\]
where we are assuming \eqref{23},  leads us to the following equations for $v$ and $u$
\begin{equation}\label{37}
\partial_t
\left(
\begin{array}{c}
v\\\\
u
\end{array}
\right) =
\left(
\begin{array}{cc}
2\partial_x& \partial_{xx}+\partial_x(v\cdot)\\\\
\partial_{xx}-v\partial_x& 2u_0\partial_x+u_{0,x}
\end{array}
\right)
\left(
\begin{array}{c}
a\\\\
b
\end{array}
\right)
\end{equation}

Thus we consider a  Lenard relation
\[
\left(
\begin{array}{cc}
-2\partial_x& \partial_{xx}+\partial_x(v\cdot)\\\\
-\partial_{xx}+v\partial_x& 2u_0\partial_x+u_{0,x}
\end{array}
\right)
\left(
\begin{array}{c}
R\\\\
S
\end{array}
\right)=0,
\]
which reduces to
\begin{align}\label{38}
\nonumber R&=\frac{1}{2}(S_x+v\,S),\\\\
\nonumber -\frac{1}{2}&S_{xxx}+2U\, S_x+U_x\, S=0,
\end{align}
where
\[
U:=-\frac{1}{2}v_x+\frac{1}{4}v^2+u.
\]
The first equation of \eqref{38} is satisfied if
\[
\left(
\begin{array}{c}
R\\\\
S
\end{array}
\right)=
\left(
\begin{array}{c}
\frac{\delta C}{\delta v}\\\\
\frac{\delta C}{\delta u}
\end{array}
\right),
\]
with
\[
C:=C[U]=\oint_{\gamma} \frac{\d z}{2\pi i}\int \d x \,\chi(U),
\]
being a functional depending on $U$. As for the second equation
of \eqref{38}, it reduces to
\[
\chi_x+\chi^2-U=0,
\]
which in turn is satisfied by
\[
\chi=\sigma-\frac{1}{2}v,
\]
with $\sigma$ being a solution of
\[
\sigma_x+\sigma^2-v\,\sigma-u=0,
\]
The last equation is verified by
\[
\sigma:=\partial_x \ln \,\psi.
\]

Therefore, we have found a quantum counterpart of the hierarchy
\eqref{25}
\begin{equation}\label{39}
 \partial_{t_N} \bu =\mathcal{J}\Big(\nabla H_N\Big)_{+}, \quad N\geq 0,
\end{equation}
where
\begin{equation}\label{38a}
\mathcal{J}:=\left(
\begin{array}{cc}
-2\partial_x& \partial_{xx}+\partial_x(v\cdot)\\\\
-\partial_{xx}+v\partial_x& 2u\partial_x+u_x
\end{array}
\right)
\end{equation}
and
\[
H_N[v,u]:=
\oint_{\gamma} \frac{\d z}{2\pi i}\int \d x \,z^N\, \chi(U).
\]
We notice that
\[
H_N[v,u]=
\oint_{\gamma} \frac{\d z}{2\pi i}\int \d x\, \frac{z^N}{2}(\sigma_+-\sigma_-),
\]
where
\[
\sigma_{\pm}=\frac{1}{2}v\pm \chi,
\]
verify
\[
L\,\psi=(\partial_x -\sigma_-)(\partial_x-\sigma_+).
\]

%
%

\noindent {\bf Acknowledgements}
\vspace{0.3cm}

L. Martinez Alonso wishes to thank the members of the
Physics Department of Lecce  University  for their warm
hospitality.

\end{document}